\documentclass[journal]{IEEEtran}
\usepackage{amsmath}
\usepackage{amsfonts}
\usepackage{amssymb}
\usepackage{algorithmic}
\usepackage{array}
\usepackage{url}
\usepackage{microtype}
\usepackage{graphicx}
\usepackage{subfigure}
\usepackage{booktabs} 
\usepackage{tabularx, enumitem}
\usepackage{hyperref}
\usepackage{cite}

\usepackage[ruled, vlined]{algorithm2e}
\DontPrintSemicolon
\usepackage{multirow}
\usepackage{setspace}

\newcommand{\rgbsymbol}{\includegraphics[height=1.5ex]{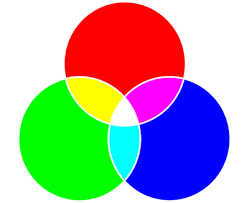}}
\begin{document}

\title{DuRIN: A Deep-unfolded Sparse Seismic Reflectivity Inversion Network}

\author{Swapnil~Mache$^\dagger$,
        Praveen~Kumar~Pokala$^\dagger$,
        Kusala~Rajendran,
        and~Chandra~Sekhar~Seelamantula,~\IEEEmembership{Senior Member,~IEEE}
\thanks{$^\dagger$ Equal contribution.}%
\thanks{Swapnil Mache is with the Centre of Excellence in Advanced Mechanics of Materials, Indian Institute of Science (IISc), Bangalore, India, and with the Department of Electrical Engineering, IISc. He was formerly with the Centre for Earth Sciences, IISc. Email:~machesanjay@iisc.ac.in.}%
\thanks{Praveen Kumar Pokala and Chandra Sekhar Seelamantula are with the Department of Electrical Engineering, IISc. E-mail:~praveenkumar.pokala@gmail.com, chandra.sekhar@ieee.org. }%
\thanks{Kusala Rajendran is with the Centre of Excellence in Advanced Mechanics of Materials, IISc. She is Retired Professor from the Centre for Earth Sciences, IISc. Email:~kusalaraj@gmail.com.}
}

\maketitle

\begin{abstract}
    We consider the reflection seismology problem of recovering the locations of interfaces and the amplitudes of reflection coefficients from seismic data, which are vital for estimating the subsurface structure. The reflectivity inversion problem is typically solved using greedy algorithms and iterative techniques. Sparse Bayesian learning framework, and more recently, deep learning techniques have shown the potential of data-driven approaches to solve the problem. In this paper, we propose a weighted minimax-concave penalty-regularized reflectivity inversion formulation and solve it through a model-based neural network. The network is referred to as deep-unfolded reflectivity inversion network (DuRIN). We demonstrate the efficacy of the proposed approach over the benchmark techniques by testing on synthetic 1-D seismic traces and 2-D wedge models and validation with the simulated 2-D Marmousi2 model and real data from the Penobscot 3D survey off the coast of Nova Scotia, Canada.
\end{abstract}


\IEEEpeerreviewmaketitle

\section{Introduction}
\begin{figure*}[!t]
  \centering
  \resizebox{0.9\linewidth}{!}{
  \includegraphics{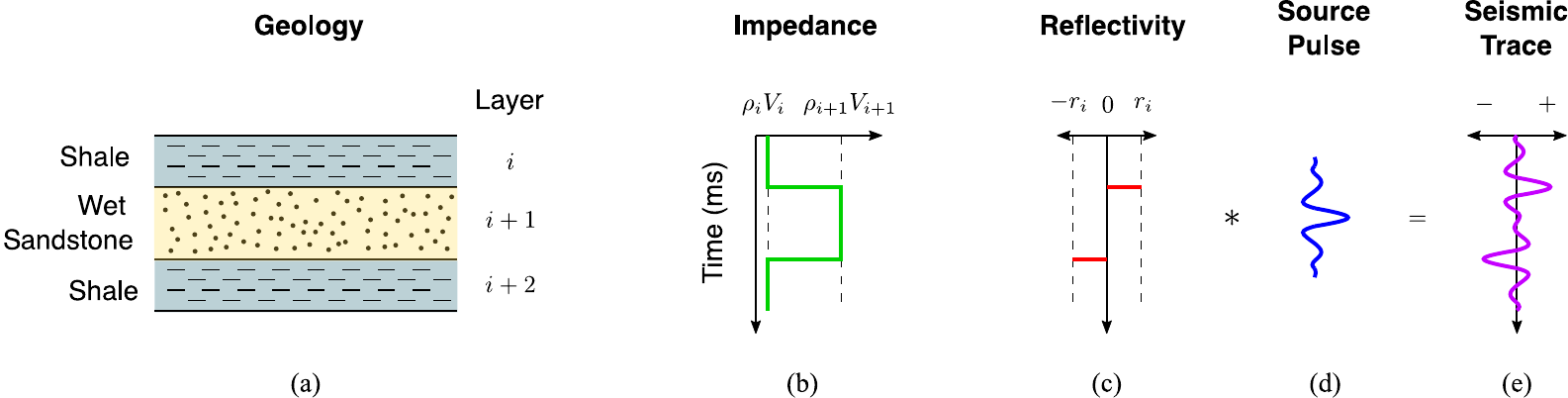}}
  \caption{\protect\rgbsymbol~An ideal three-layer subsurface model. The operator $*$ denotes convolution.}\label{fig:model}
\end{figure*}

\IEEEPARstart{R}{eflectivity} inversion is an important deconvolution problem in reflection seismology, helpful in characterizing the layered subsurface structure. The subsurface is modeled as having sparse reflectivity localized at the interfaces between two layers, assuming largely horizontal and parallel layers, each with constant impedance \cite{oldenburg1983recovery, yilmaz2001seismic}, or in other words, a piecewise-constant impedance structure. Figure \ref{fig:model} shows a $3$-layer model of the subsurface, with a wet sandstone between two shale layers \cite{russell2019machine}. 
The reflection coefficients ($r_{i}$) (Fig.~\ref{fig:model}(c)) at the interface between two adjoining layers $i$ and $i+1$ (Fig.~\ref{fig:model}(a)) are related to the subsurface geology \cite{oldenburg1983recovery} through the relation:
\begin{equation}\label{rhov}
    r_{i} = \frac{\rho_{i+1}V_{i+1} - \rho_{i}V_{i}}{\rho_{i+1}V_{i+1} + \rho_{i}V_{i}},
\end{equation}%
where $\rho_i$ and $V_i$ are the density and P-wave velocity, respectively, of the $i^{th}$ layer. The product of density ($\rho$) and P-wave velocity ($V$) is known as the acoustic impedance (Fig.~\ref{fig:model}(b)).

Such a layered subsurface is often recovered through the widely used convolutional model \eqref{convmodel}, wherein the observed seismic data is modeled as the convolution between the source pulse ${\boldsymbol{h}}$ (Fig.~\ref{fig:model}(d)) and the Earth response or reflectivity ${\boldsymbol{x}} \in \mathbb{R}^{n}$ (Fig.~\ref{fig:model}(c)) \cite{taylor1979deconvolution, shearer2009introduction}. The observation at the surface is a noisy seismic trace ${\boldsymbol{y}} \in \mathbb{R}^{n}$ (Fig.~\ref{fig:model}(e)) given by:
\begin{equation}\label{convmodel}
    {\boldsymbol{y}} = {\boldsymbol{h}}*{\boldsymbol{x}} + {\boldsymbol{n}},
\end{equation}%
where ${\boldsymbol{n}}$ is the measurement noise, $*$ denotes convolution, and ${\boldsymbol{h}}$ is assumed to be a Ricker wavelet. When the wavelet is known, the linear inverse problem \eqref{convmodel} can be solved within the sparsity framework by modeling the reflectivity ${\boldsymbol{x}}$ as sparse, as the significant interfaces are important in the inversion. Further, the location of the interfaces (in other words, the support recovery) is prioritized over amplitude recovery \cite{shearer2009introduction}.

\subsection{Prior Art}
We broadly classify the prior work related to reflectivity inversion into two categories, namely, optimization-based techniques and data-driven techniques.

\subsubsection{Optimization Techniques}
The solution $\boldsymbol{x}$ in \eqref{convmodel} can be recovered by minimizing the classical least-squares (LS) objective function:%
\begin{equation}\label{leastsquares}
    \arg\min_{{\boldsymbol{x}}}\frac{1}{2}{\left\| {\boldsymbol{h}}*{\boldsymbol{x}}-{\boldsymbol{y}} \right\|}_{2}^{2}.
\end{equation}%
However, the finite length and measurement noise, and the loss of low and high-frequency information due to convolution of the reflectivity with a bandlimited wavelet \cite{oldenburg1983recovery, berkhout1977least, debeye1990lp}, lead to non-unique solutions to the above problem \cite{yuan2019seismic}. This ill-posed inverse problem is tackled by employing a sparsity prior on the solution \cite{tarantola2005inverse}. The $\ell_1$-norm regularization is widely preferred by virtue of its convexity \cite{oldenburg1983recovery, taylor1979deconvolution, liu2014fast, zhang2017inversion}. The $\ell_1$-regularized reflectivity inversion problem is posed as follows:
\begin{equation}\label{l1}
    \arg\min_{{\boldsymbol{x}}} \, \frac{1}{2}{\left\| {\boldsymbol{h}}*{\boldsymbol{x}}-{\boldsymbol{y}} \right\|}_{2}^{2} + \lambda {\left\| {\boldsymbol{x}} \right\|}_{1},
\end{equation}%
where $\lambda$ is the regularization parameter.

Zhang and Castagna \cite{zhang2011seismic} adopted basis-pursuit inversion (BPI) to solve the problem in \eqref{l1}, based on the basis-pursuit de-noising algorithm proposed by \cite{chen2001atomic}. Algorithms such as the Iterative Shrinkage-Thresholding Algorithm (ISTA) \cite{daubechies2004iterative} and its faster variant, FISTA \cite{beck2009fast}, have been proposed to solve the $\ell_1$-norm regularized deconvolution problem. Relevant to the problem under consideration, \cite{perez2012inversion} adopted FISTA for reflectivity inversion. Further studies by \cite{perez2013high} and \cite{li2020debiasing} adopted FISTA along with debiasing steps of LS inversion and ``adding back the residual'', respectively, to improve amplitude recovery after support estimation using FISTA. 

The sparsity of seismic data is difficult to estimate in practice and the adequacy of the $\ell_1$ norm for the reflectivity inversion problem has not been established \cite{yuan2019seismic, li2019optimal}. Also, $\ell_1$-norm regularization results in a biased estimate of ${\boldsymbol{x}}$ \cite{candes2008enhancing, zhang2010analysis, selesnick2017sparse}. In their application of FISTA to the seismic reflectivity inversion problem, \cite{perez2013high} and \cite{li2020debiasing} observed an attenuation of reflection coefficient magnitudes. They adopted post-processing debiasing steps \cite{wright2009sparse} to tackle the bias introduced due to the $\ell_1$-norm regularization. Nonconvex regularizers such as the minimax concave penalty (MCP) \cite{zhang2010nearly} have been shown to overcome the shortcomings of $\ell_1$ regularization in inverse problems. The advantages of adopting nonconvex regularizers over $\ell_1$ have been demonstrated in sparse recovery problems \cite{selesnick2017sparse, zhang2010nearly, woodworth2016compressed}. Particularly pertinent to this discussion is the data-driven $\ell_{q}$-norm regularization $(0<q<1)$ for seismic reflectivity inversion proposed by \cite{li2019optimal}, wherein the optimal $q$ was chosen based on the input data. 

\subsubsection{Data-driven Methods}
Recent works have employed data-driven approaches for solving inverse problems in seismology and geophysics \cite{yuan2013spectral, yuan2019seismic, russell2019machine, bergen2019machine, adler2021deep, lewis2017deep, richardson2018seismic, ovcharenko2019deep, sun2020extrapolated, araya2018deep, kim2018geophysical, wang2018velocity, wu2018inversionnet, yang2019deep, adler2019deep, das2019convolutional, park2020automatic, wu2020seismic, wang2020velocity, biswas2019prestack, alfarraj2019semi, araya2019deep, wang2019seismic, mosser2020stochastic, zhang2020data}. Further, learning-based approaches have also been explored for seismic reflectivity inversion (or {\it reflectivity model building}) \cite{russell2019machine, adler2021deep, kim2018geophysical, biswas2019prestack}. Bergen {\it et al.} \cite{bergen2019machine} underscored the importance of leveraging the data and model-driven aspects of inverse problems in solid Earth geoscience while incorporating machine learning (ML) into the loop. In a comprehensive review of deep learning approaches for inverse problems in seismology and geophysics, Adler {\it et al.} \cite{adler2021deep} assessed learning-based approaches that enhanced the performance of full waveform inversion \cite{lewis2017deep, richardson2018seismic, ovcharenko2019deep, sun2020extrapolated} and end-to-end seismic inversion \cite{araya2018deep, kim2018geophysical, wang2018velocity, wu2018inversionnet, yang2019deep, adler2019deep, das2019convolutional, park2020automatic, wu2020seismic, wang2020velocity}. They also reviewed recent solutions based on physics-guided architectures \cite{biswas2019prestack, alfarraj2019semi} and deep generative modeling \cite{araya2019deep, wang2019seismic, mosser2020stochastic, zhang2020data} that are still in a nascent stage as it pertains to seismic inversion.

The application of sparse Bayesian learning (SBL) for reflectivity inversion has also been explored \cite{yuan2013spectral, yuan2019seismic}, where the sparse reflection coefficients are recovered by maximizing the marginal likelihood. This is done through a sequential algorithm-based approach (SBL-SA) to update the sparsity-controlling hyperparameters \cite{yuan2013spectral}, or through the expectation-maximization algorithm (SBL-EM) \cite{yuan2019seismic, wipf2004sparse}. The application of neural networks to inversion problems in geophysics \cite{adler2021deep}, particularly reflectivity inversion, is a recent development \cite{russell2019machine, kim2018geophysical}. Kim and Nakata \cite{kim2018geophysical} used an elementary feedforward neural network to recover the sparse reflectivity. They obtained superior support recovery but low amplitude recovery using the neural network compared to a least-squares approach. Further, \cite{russell2019machine} discussed the suitability of machine learning approaches such as feedforward neural networks \cite{kim2018geophysical}. He observed that data-driven techniques outperform conventional deconvolution approaches when knowledge about the underlying geology is limited while also providing a computational advantage. Although, in applications pertaining to geoscience, and specifically, geophysics, the level of model interpretability is critical as one aims to gain physical insights into the system under consideration \cite{bergen2019machine}. Insights into an inverse problem, in addition to savings in terms of computational times, can be gained through deep neural network architectures that are informed by the sparse linear inverse problem itself \cite{bergen2019machine}. 

To that end, \cite{gregor2010learning} proposed a new class of data-driven framework architectures based on unfolding iterative algorithms into neural network architectures \cite{monga2021algorithm}. They proposed learned ISTA (LISTA), based on unrolling the update steps of ISTA \cite{daubechies2004iterative} into layers of a feedforward neural network. Subsequent studies have demonstrated the efficacy of this class of model-based deep learning architectures in compressed sensing and sparse-recovery applications \cite{mahapatra2017deep, mukherjee2017dnns, zhang2017ista, borgerding2017amp, sreter2018learned, liu2019deep, pokala2019firmnet, li2020efficient, shlezinger2020modelbased, pokala2020confirmnet, wang2020dnu, tolooshams2020deep, jawali2020cornet, tolooshams2021unfolding}. Monga {\it et al.} \cite{monga2021algorithm} provided a review on algorithm unrolling in the context of imaging, vision and recognition, speech processing, and other signal and image processing problems. Deep-unfolding combines the advantages of both data-driven and iterative techniques. 

\subsection{Motivation and Contribution}

The limitations of $\ell_1$ regularization \cite{candes2008enhancing, zhang2010analysis, selesnick2017sparse}, in addition to the challenge of estimating the sparsity of seismic reflection data \cite{yuan2019seismic, li2019optimal}, can be overcome through nonconvex regularization \cite{li2019optimal}. However, the corresponding nonconvex cost suffers from local minima, and optimization becomes quite challenging. The hyper-parameters of most iterative techniques are set heuristically, which is likely to yield suboptimal solutions. Machine learning approaches outperform conventional techniques in reflectivity inversion \cite{russell2019machine}, especially in support recovery \cite{kim2018geophysical}, which is given higher priority \cite{shearer2009introduction}. Deep-unrolled architectures \cite{monga2021algorithm}, which belong to a class of model-based neural networks, are not entirely dissociated from the underlying physics of the problem, which is a potential risk in elementary data-driven approaches such as feedforward neural networks \cite{russell2019machine, kim2018geophysical, bergen2019machine}. These factors motivate us to employ a data-driven nonconvex regularization strategy and solve the reflectivity inversion problem through deep-unrolled architectures. 

The contributions of this paper are summarized as follows. 
\begin{enumerate}
    \item We construct an optimization cost for sparse seismic reflectivity inversion based on the weighted counterpart of the {\it minimax-concave} penalty (MCP) \cite{zhang2010nearly}, where each component of the sparse reflectivity ${\boldsymbol{x}}$ is associated with a different weight. 
    \item The resulting optimization algorithm, the iterative firm-thresholding algorithm (IFTA) \cite{pokala2019firmnet} is unfolded into the deep-unfolded reflectivity inversion network (DuRIN). To the best of our knowledge, model-based architectures have not been explored for solving the seismic reflectivity inversion problem. In fact, deep-unrolling has not been employed for solving seismic inverse problems \cite{adler2021deep}. 
    \item The efficacy of the proposed formulation with respect to the state of the art is demonstrated over synthetic $1$-D and $2$-D data and on simulated as well as real datasets. 
\end{enumerate}
%

\section{Weighted-MCP-Regularized Reflectivity Inversion}

In this study, we use the weighted counterpart of the nonconvex minimax-concave penalty (MCP) \cite{zhang2010nearly} defined as:
\begin{equation}
    g_{ {\boldsymbol{\gamma}} }( {\boldsymbol{x}}; {\boldsymbol{\mu}}) = \sum\limits_{i=1}^n g_{\gamma_i}(x_i; \mu_i),
\end{equation}
where ${\boldsymbol{x}} = \left[ x_1, x_2, x_3, \ldots, x_n \right]$, ${\boldsymbol{\mu}} = \left[ \mu_1, \mu_2, \mu_3, \ldots, \mu_n \right]$, ${\boldsymbol{\gamma}} = \left[ \gamma_1, \gamma_2, \gamma_3, \ldots, \gamma_n \right]$, and
\begin{equation}\label{MCP}
    g_{\gamma_i}(x_i;\mu_i) = \begin{cases}{\mu_i\left|x_i\right| - \cfrac{{\left| x_i\right|}^2}{2\gamma_i},}
    \quad {\text{for} \left| {x}_i \right| \leq {\gamma}_i ,} \\ 
    {\cfrac{{{{\mu}}_i }^{2}{\gamma}_i }{2},}
    \quad\quad\qquad\qquad {\text{for} \left| {x}_i \right| \geq {\gamma}_i {{\mu}}_i ,} \end{cases}
\end{equation}
$\forall~i \in [\![1,n]\!]$. The trainable parameters of $g_{ {\boldsymbol{\gamma}} }( {\boldsymbol{x}}; {\boldsymbol{\mu}})$, ${\boldsymbol{\mu}} \in \mathbb{R}^n_{>0}$ and ${\boldsymbol{\gamma}} \in \mathbb{R}^n_{>1}$, are learned in a data-driven setting. $\mathbb{R}^n_{>t}$ denotes the set of vectors in $\mathbb{R}^n$ with entries greater than $t$.
\begin{figure}
    \centering
    \includegraphics[width=0.8\linewidth]{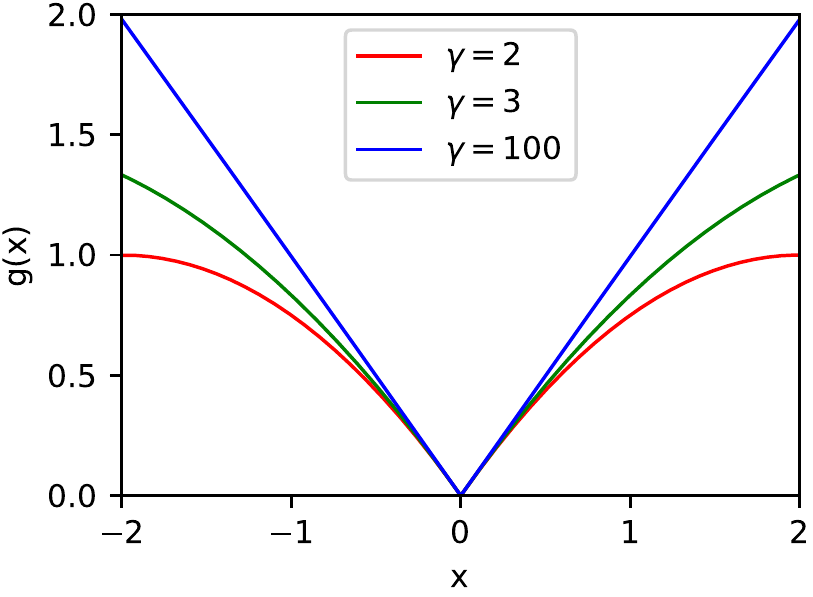}
    \caption{\protect\rgbsymbol~The minimax-concave penalty corresponding to $\gamma=2$, $\gamma=3$ and $\gamma=100$.}
    \label{fig:penalty_functions}
\end{figure}

\begin{figure}
    \centering
    \includegraphics[width=0.8\linewidth]{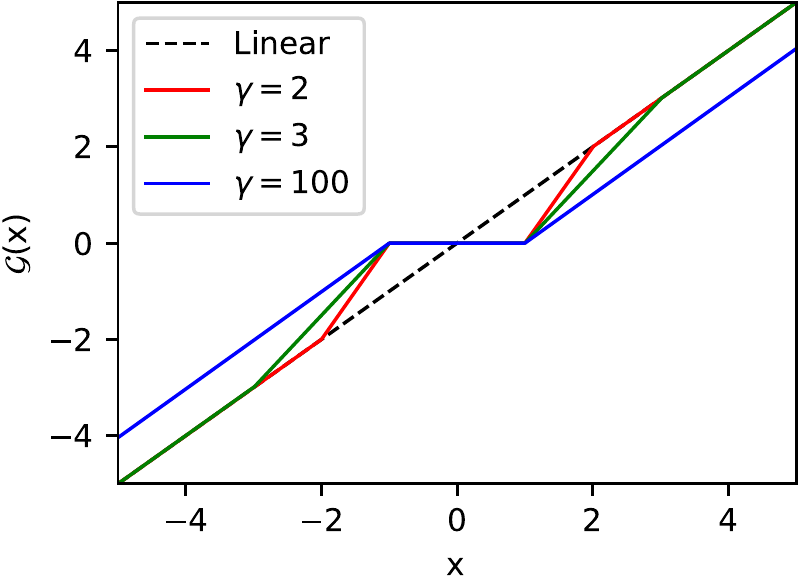}
    \caption{\protect\rgbsymbol~The proximal operators corresponding to the MCP for $\gamma=2$, $\gamma=3$ and $\gamma=100$.}
    \label{fig:prox_operators}
\end{figure}

\subsection{Problem Formulation}
Formulating the optimization problem using the weighted MCP regularizer defined above leads to the objective function:%
\begin{equation}\label{regularization_problem}
    \arg\min_{{\boldsymbol{x}}} \, \Big\{ {J({\boldsymbol{x}})} =  \underbrace{\frac{1}{2}{\left\| {\boldsymbol{h}}*{\boldsymbol{x}}-{\boldsymbol{y}} \right\|}_{2}^{2}}_{F({\boldsymbol{x}})} + g_{{\boldsymbol{\gamma}}}( {\boldsymbol{x}}; {\boldsymbol{\mu}}) \Big\}.
\end{equation}%
The objective function ${J}$, data-fidelity term $F$, and the regularizer $g$ satisfy the following properties, which are essential for minimizing ${J}$:
\begin{enumerate}
	\item[P1.] $F : \mathbb{R}^{n} \rightarrow \left (- \infty,\infty \right] $ is proper, closed, and $L$-smooth, i.e., $\Vert \nabla F({\boldsymbol{x}}) - \nabla F({\boldsymbol{y}})\Vert_{2} \leq L \Vert {\boldsymbol{x}} - {\boldsymbol{y}} \Vert_{2}$.
	\item[P2.] $g$ is lower semi-continuous.
	\item[P3.] ${J}$ is bounded from below, i.e., inf ${J} > -\infty$.
\end{enumerate}
Next, we optimize the problem stated in \eqref{regularization_problem} through the Majorization-Minimization (MM) approach \cite{figueiredo2007majorization}, and the resulting algorithm, the iterative firm-thresholding algorithm (IFTA) \cite{pokala2019firmnet}. Unfolding the iterations of IFTA results in a learnable network called deep-unfolded reflectivity inversion network (DuRIN), which has a similar architecture to FirmNet \cite{pokala2019firmnet}.

\subsection{IFTA: Iterative firm-thresholding algorithm}

Due to P1, there exists $\eta < 1/L$ such that $F({\boldsymbol{x}})$ is upper-bounded locally by a quadratic expansion about ${\boldsymbol{x}} = {\boldsymbol{x}}^{(k)}$ as:
\begin{equation}
    F({\boldsymbol{x}}) \leq {Q^{(k)} \left( \boldsymbol{x} \right)},
\end{equation}
where ${Q^{(k)} \left( \boldsymbol{x} \right)} = F({\boldsymbol{x}}^{(k)}) + ({\boldsymbol{x}} - {\boldsymbol{x}}^{(k)})^{\intercal}\nabla F({\boldsymbol{x}}^{(k)}) + \frac{1}{2 \eta} {\left\| {\boldsymbol{x}} - {\boldsymbol{x}}^{(k)} \right\|}_{2}^{2}$. The majorizer to the objective function $J$ at ${\boldsymbol{x}}^{(k)}$ is given by:
\begin{equation}\label{majorizer}
    H^{(k)}({\boldsymbol{x}}) = Q^{(k)} \left( \boldsymbol{x} \right) + g_{{\boldsymbol{\gamma}}}( {\boldsymbol{x}}; {\boldsymbol{\mu}}),
\end{equation}
such that $H^{(k)}({\boldsymbol{x}})\geq J(\boldsymbol{x})$. The update equation for minimizing $H^{(k)}({\boldsymbol{x}})$ to obtain ${\boldsymbol{x}}^{(k+1)}$ is given by,%
\begin{align}
    {\boldsymbol{x}}^{(k+1)} = \arg\min_{{\boldsymbol{x}}} \frac{1}{2} {\left\| {\boldsymbol{x}}^{(k)} + \eta{\boldsymbol{h}}'* ({\boldsymbol{y}}-{\boldsymbol{h}}*{\boldsymbol{x}}^{(k)}) - {\boldsymbol{x}} \right\|}_{2}^{2} \nonumber \\
    + \eta g_{{\boldsymbol{\gamma}}}( {\boldsymbol{x}}; {\boldsymbol{\mu}}).
\end{align}%

The proximal operator of the penalty function $g_{{\gamma}_i}( {x}_i; {\mu}_i)$ is the firm-thresholding operator \cite{voronin2013new} given by
\begin{align}
    \mathcal{G}_{{\gamma}_i}^{g}({x}_i; {\mu}_i) = \begin{cases} {0,}
    \qquad\qquad\quad\, {\text{for} \; \left| {x}_i \right| \leq {{\mu}}_i,} \\[5pt]
    {\text{sgn} \left({x}_i\right) \odot \cfrac{{\gamma}_i}{{\gamma}_i - 1} \left( \left| {x}_i \right| - {{\mu}}_i \right),}& \\[8pt] 
    \qquad\qquad\qquad 
    {\text{for} \; {{\mu}}_i < \left| {x}_i \right| \leq {\gamma}_i {{\mu}}_i,} \\
    {{x}_i,}
    \qquad\qquad\quad{\text{for} \; \left| {x}_i \right| > {\gamma}_i {{\mu}}_i}. \end{cases}
\label{firm_thresh}
\end{align}

The update step at the $(k+1)^{th}$ iteration is given by
\begin{equation}
    \begin{split}
        {\boldsymbol{x}}^{(k+1)} = \mathcal{G}_{{\boldsymbol{\gamma}}}^{g} \left( {\boldsymbol{x}}^{(k)} + \frac{\eta}{2} {\boldsymbol{h}}' * \left( {\boldsymbol{y}} - {\boldsymbol{h}} * {\boldsymbol{x}}^{(k)} \right) \right),
    \end{split}
    \label{ifta}
\end{equation}%
where ${\boldsymbol{h}}'$ is the flipped version of ${\boldsymbol{h}}$. Algorithm \eqref{alg:ifta} lists the steps involved in solving the optimization problem \eqref{regularization_problem}. The update in \eqref{ifta} involves convolutions followed by nonlinear activation (the firm threshold in this case), and can therefore be represented as a layer in a neural network.
\begin{algorithm}[t]
    \caption{IFTA: Iterative Firm-Thresholding Algorithm}\label{alg:ifta}
    \KwIn{$\eta = 1/{\left\| {\boldsymbol{h}} \right\|}_2^2,~ k_{max}$, data ${\boldsymbol{y}}$, wavelet ${\boldsymbol{h}}$, learnable parameters ${\boldsymbol{\mu}}$, ${\boldsymbol{\gamma}}$}
    Initialization: $ {\boldsymbol{x}}^{(0)} = 0 $\;
    \While{$k < k_{max}$}{
    $\boldsymbol{z}^{(k+1)} = {\boldsymbol{x}}^{(k)} - \frac{\eta}{2}\left( {\boldsymbol{h}}'* ({\boldsymbol{y}} - {\boldsymbol{h}}*{{\boldsymbol{x}}^{(k)}}) \right)$\;
    ${\boldsymbol{x}}^{(k+1)} = \mathcal{G}_{{\boldsymbol{\gamma}}}^{g} \left( \boldsymbol{z}^{(k+1)}; {\boldsymbol{\mu}} \right)$\;
    $k = k + 1$\;
    }
    \KwOut{$\hat{{\boldsymbol{x}}} \leftarrow {\boldsymbol{x}}^{(k)}$}
\end{algorithm}%

\subsection{DuRIN: Deep-unfolded Reflectivity Inversion Network}
\label{subsec:durin}%
As mentioned in the previous section, the update in IFTA \eqref{ifta} can be interpreted as a layer in a neural network, and hence, we unroll the iterations of the algorithm into layers of a neural network to solve the problem given in \eqref{regularization_problem}. The proposed deep-unrolled architecture for sparse seismic reflectivity inversion is called Deep-unfolded Reflectivity Inversion Network (DuRIN), which resembles FirmNet \cite{pokala2019firmnet}. The input-output for each layer in DuRIN is given by
\begin{equation}
    {\boldsymbol{x}}^{(k+1)} = \mathcal{G}_{{\boldsymbol{\gamma}}}^{g} ( \mathbf{W} {\boldsymbol{y}} + \mathbf{S} {\boldsymbol{x}}^{(k)}),
    \label{eq:durin}
\end{equation}
where \( \mathbf{W}=\eta\mathop{{\boldsymbol{h}}'} \) and \( \mathbf{S}=\mathbf{I}-\eta\mathop{{\boldsymbol{h}}' {\boldsymbol{h}}} \) \cite{gregor2010learning} are initialized as Toeplitz matrices, and are dense and unstructured in the learning stage. The parameters that need to be learned are the matrices $\mathbf{W}$ and $\mathbf{S}$ and the thresholds (${\boldsymbol{\mu}}, {\boldsymbol{\gamma}}$), subject to ${\boldsymbol{\mu}}>{\boldsymbol{0}}, {\boldsymbol{\gamma}}>{\boldsymbol{1}}$, where ${\boldsymbol{0}}$ is the null vector and ${\boldsymbol{1}}$ is the vector of all ones. For training, we employ the smooth $\ell_1$ loss $\left( \text{Smooth}~\ell_1 = \beta \left\|{{\boldsymbol{x}} - \hat{{\boldsymbol{x}}}} \right\|_{1} + (1-\beta) \left\|{{\boldsymbol{x}} - \hat{{\boldsymbol{x}}}} \right\|_{2}^{2},~0<\beta \leq 1 \right)$ as the training cost, computed between the true reflectivity ${\boldsymbol{x}}$ and prediction $\hat{{\boldsymbol{x}}}$. In our experimental setup, $\beta=1$, and it is not a trainable parameter.

\begin{figure*}[t]
    \centering
    \includegraphics[width=0.8\linewidth]{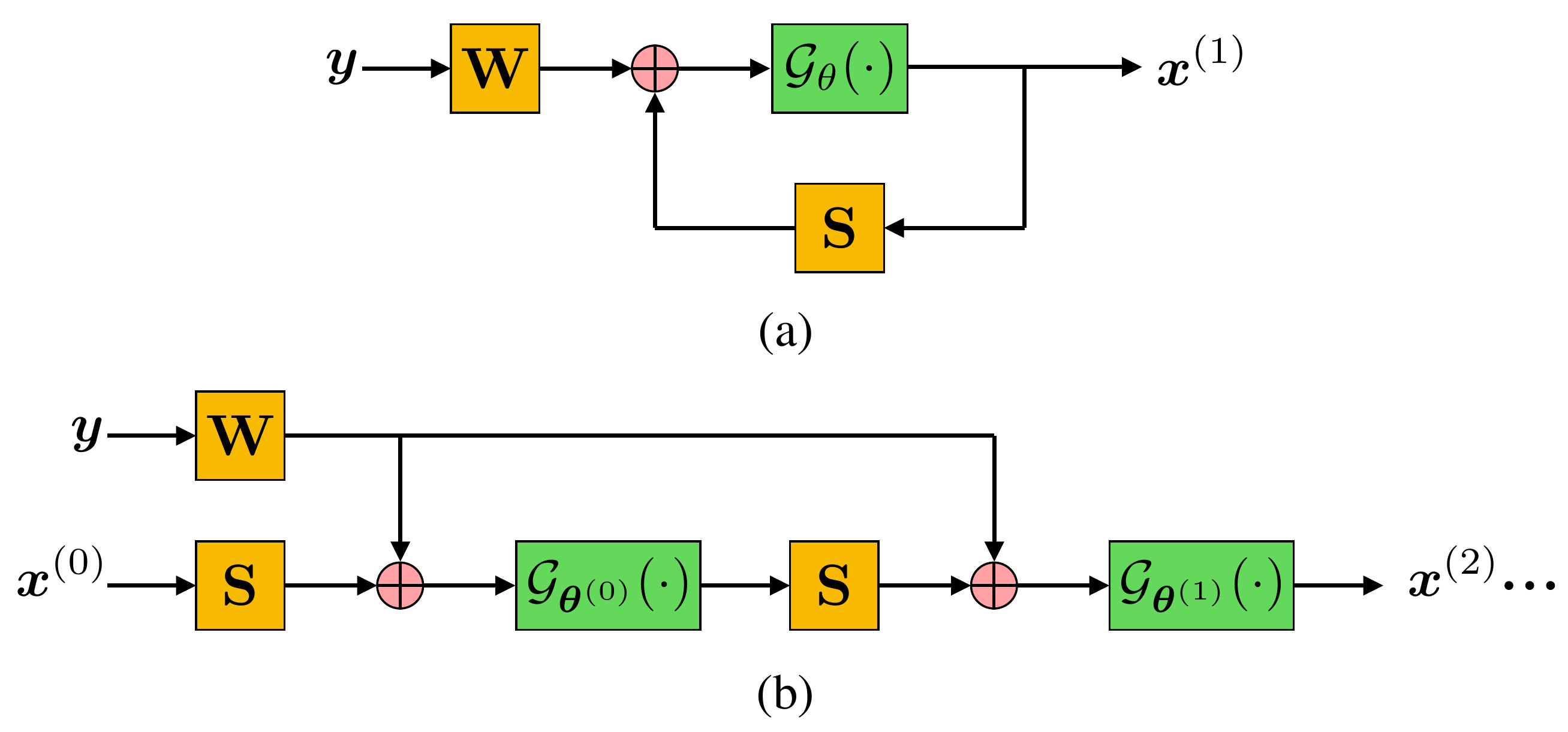}
    \caption{\protect\rgbsymbol~Schematic for (a) one iteration of the iterative firm-thresholding algorithm (IFTA) \cite{pokala2019firmnet}, and (b) 2 layers of the deep-unfolded reflectivity inversion network (DuRIN), which corresponds to FirmNet \cite{pokala2019firmnet}. A layer in the network (b) corresponds to an iteration in (a). The operator $\oplus$ represents element-wise sum.} \label{fig:schematic}
\end{figure*}

To improve amplitude recovery of DuRIN during inference, we re-estimate the amplitudes of the estimated sparse vector $\hat{{\boldsymbol{x}}}$ over the supports given by DuRIN as: $\hat{{\boldsymbol{x}}}_{\hat{\mathcal{S}}_{\hat{\boldsymbol{x}}}} = \mathbf{H}_{\hat{\mathcal{S}}_{\hat{\boldsymbol{x}}}}^{\dagger}{\boldsymbol{y}}$, where $\hat{\mathcal{S}}_{\hat{\boldsymbol{x}}}$ is the support, or the non-zero locations, of $\hat{\boldsymbol{x}}$, and $\mathbf{H}_{\hat{\mathcal{S}}_{\hat{\boldsymbol{x}}}}^{\dagger}$ is the pseudo-inverse of the Toeplitz matrix $\mathbf{H}$ of the kernel ${\boldsymbol{h}}$ over the support $\hat{\mathcal{S}}_{\hat{\boldsymbol{x}}}$. The steps of DuRIN for $k_{max}$ layers are listed in Algorithm~\ref{alg:durin}. Figure \ref{fig:schematic} illustrates a 3-layer architecture for DuRIN.%

%

%
We consider two variants of DuRIN, depending on the value of $\gamma_i,~\forall~i \in [\![1,n]\!]$. The generalized variant of DuRIN based on IFTA \cite{pokala2019firmnet}, where ${\gamma}_i$ are different across components and layers, is referred to as DuRIN-$1$, and corresponds to FirmNet \cite{pokala2019firmnet}. As $\gamma_i \to \infty$, the firm-thresholding function approaches the soft thresholding function \cite{selesnick2017sparse}. We call this variant DuRIN-$2$, which resembles LISTA \cite{gregor2010learning}.

\begin{algorithm}[t]
\caption{DuRIN: Deep-unfolded Reflectivity Inversion Network}\label{alg:durin}
\KwIn{$ {\boldsymbol{y}},~ \eta = 1/{\left\| \mathbf{H} \right\|}_2^2,~ k_{max},~ {\boldsymbol{\mu}},~ {\boldsymbol{\gamma}}$}
Initialization: $\mathbf{W},~ \mathbf{S},~ {\boldsymbol{x}}^{(0)} = \mathcal{G}_{{\boldsymbol{\gamma}}}^{g} (\mathbf{W}{{\boldsymbol{y}}}) $\;
\While{$k < k_{max}$}{
$\mathbf{c}^{(k+1)} = \mathbf{W}{{\boldsymbol{y}}} + \mathbf{S}{{\boldsymbol{x}}^{(k)}} $\;
${\boldsymbol{x}}^{(k+1)} = \mathcal{G}_{{\boldsymbol{\gamma}}}^{g} (\mathbf{c}^{(k+1)}) $\;
$k = k + 1$\;
}
$\hat{{\boldsymbol{x}}}_{\hat{\mathcal{S}}_{\hat{\boldsymbol{x}}}} = \mathbf{H}_{\hat{\mathcal{S}}_{\hat{\boldsymbol{x}}}}^{\dagger}{\boldsymbol{y}}$\;
\KwOut{$\hat{{\boldsymbol{x}}} \leftarrow {\boldsymbol{x}}^{(k)}$}
\end{algorithm}%

\section{Experimental Results}
\label{sec:experiments}
We present experimental results for the proposed networks, namely, DuRIN-$1$ and DuRIN-$2$, on synthetic $1$-D traces and $2$-D wedge models \cite{hamlyn2014thin}, the simulated $2$-D Marmousi$2$ model \cite{martin2006marmousi2}, and real $3$-D field data from the Penobscot $3$D survey off the coast of Nova Scotia, Canada \cite{penobscot3d}. We evaluate the performance of DuRIN-$1$ and DuRIN-$2$ in comparison with the benchmark techniques, namely, basis-pursuit inversion (BPI) \cite{chen2001atomic, zhang2011seismic}, fast iterative shrinkage-thresholding algorithm (FISTA) \cite{beck2009fast, perez2012inversion}, and expectation-maximization-based sparse Bayesian learning (SBL-EM) \cite{wipf2004sparse, yuan2019seismic}. To quantify the performance, we employ the objective metrics listed in the following section.

\subsection{Objective Metrics}
\label{subsec:metrics}
We evaluate the results using statistical parameters that measure the amplitude and support recovery between the ground-truth sparse vector ${\boldsymbol{x}}$ and the predicted sparse vector $\hat{{\boldsymbol{x}}}$.
\begin{enumerate}[wide, labelwidth=0pt, labelindent=0pt, noitemsep, nolistsep]
    \item {Correlation Coefficient (CC) \cite{freedman2007statistics}:%
    \begin{align*}
        \text{CC} = \frac{\langle {\boldsymbol{x}}, \hat{{\boldsymbol{x}}} \rangle - \langle {\boldsymbol{x}}, {{\boldsymbol{1}}} \rangle \langle \hat{{\boldsymbol{x}}}, {{\boldsymbol{1}}} \rangle}{\sqrt{\left( \left\| {{\boldsymbol{x}}} \right\|_2^2 - {\langle {\boldsymbol{x}}, {{\boldsymbol{1}}} \rangle}^2 \right) \left( \left\| \hat{{\boldsymbol{x}}} \right\|_2^2 - {\langle \hat{{\boldsymbol{x}}}, {{\boldsymbol{1}}} \rangle}^2 \right) }},
    \end{align*}%
    where $\langle \cdot, \cdot \rangle$ denotes inner product, and ${\boldsymbol{1}}$ denotes a vector of all ones.
    }%
    \item Relative Reconstruction Error (RRE) and Signal-to-Reconstruction Error Ratio (SRER), defined as follows:%
    \begin{align*}
        \text{RRE} = \frac{{\left\| \hat{{\boldsymbol{x}}} - {\boldsymbol{x}} \right\|}_2^2}{{\left\| {\boldsymbol{x}} \right\|}_2^2}, \quad \text{SRER} = 10\mathop {\log\limits_{10} \left(\frac{{\left\| {{\boldsymbol{x}}} \right\|_2^2}}{{\left\| {\hat{{\boldsymbol{x}}} - {{\boldsymbol{x}}}} \right\|_2^2}}\right){\text{dB}}}.
    \end{align*}
    \item {Probability of Error in Support (PES):%
    \begin{align*}
        \text{PES} = \left( \frac{{{\text{max}}(\left| {\hat{\mathcal{S}}_{\hat{\boldsymbol{x}}}} \right|,\left| {{\mathcal{S}}_{\boldsymbol{x}}} \right|) - \left| {\hat{\mathcal{S}}_{\hat{\boldsymbol{x}}}\mathop {\cap} {\mathcal{S}_{\boldsymbol{x}}}} \right|}}{{{\text{max}}(\left| {\hat{\mathcal{S}}_{\hat{\boldsymbol{x}}}} \right|,\left| {\mathcal{S}_{\boldsymbol{x}}} \right|)}}\right), 
    \end{align*}%
    where $|\cdot|$ denotes the cardinality of the argument, and ${\mathcal{S}}_{\boldsymbol{x}}$ and $\hat{\mathcal{S}}_{\hat{\boldsymbol{x}}}$ denote the supports of ${\boldsymbol{x}}$ and $\hat{{\boldsymbol{x}}}$, respectively.}
\end{enumerate}%

\subsection{Training Phase}\label{subsec:training_phase}
The generation or acquisition of training data appropriately representing observed seismic data is crucial for the application of a data-driven approach to the reflectivity inversion problem \cite{kim2018geophysical}. We generate synthetic training data as $5\times10^{5}$ seismic traces, each of length $300$ samples, obtained by the convolution between $1$-D reflectivity profiles and a Ricker wavelet. The reflectivity profiles consist of $200$ samples each, and are padded with zeros before convolution such that their length is equal to that of the seismic traces \cite{russell2019machine}. Amplitudes of the reflection coefficients range from $-1.0$ to $1.0$, and the sparsity factor ($k$), i.e., the ratio of the number of non-zero elements to the total number of elements in a trace, is set to $0.05$. The locations of the reflection coefficients are picked uniformly at random, with no constraint on the minimum spacing between two reflectors/spikes, or in other words, the reflectors can be away only by one sampling interval. The reflectivity values, chosen randomly from the defined range of $-1.0$ to $1.0$, are assigned to the selected locations. The sampling interval, the increment in the amplitudes of the reflection coefficients, and the wavelet frequency vary based on the dataset. 

The initial hyperparameters and the optimum number of layers in the network also vary depending on the parameters of the dataset, such as the sampling interval and the wavelet frequency. We initially consider six layers for both DuRIN-$1$ and DuRIN-$2$ for training and keep appending five layers at a time to arrive at the optimum number of layers. For the training phase, we set the batch size to $200$, use the ADAM optimizer \cite{kingma2014adam} and set the learning rate to $1\times10^{-4}$, with an input measurement signal-to-noise ratio (SNR) of $20$ dB to ensure the robustness of the proposed models against noise in the testing data \cite{kim2018geophysical}. The proposed networks, DuRIN-$1$ and DuRIN-$2$, are trained on $1$-D data, and operate on $1$-D, $2$-D, and $3$-D data trace-by-trace. In the following sections, we provide experimental results on synthetic, simulated, and real data.

\subsection{Testing Phase: Synthetic Data}\label{subsec:testing_phase}
\subsubsection{Synthetic 1-D Traces}\label{subsubsec:1d_trace}
\begin{figure*}[t]
    \centering
    \includegraphics[width=0.55\linewidth]{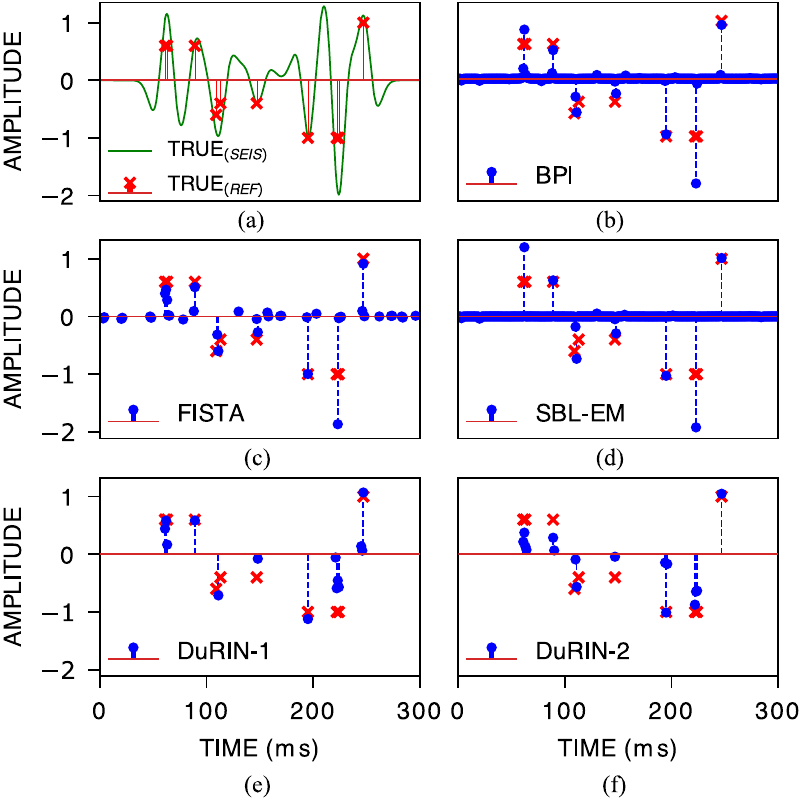}
    \caption{\protect\rgbsymbol~Sample results for recovered reflectivity from a synthetic $1$-D seismic trace out of $1000$ test realizations reported in Table \ref{table:1d_trace}. (a) The true seismic traces (TRUE$_{SEIS}$) and reflectivity (TRUE$_{REF}$); (b)-(d) Recovered $1$-D reflectivity profiles for the benchmark techniques show that they fail to distinguish between closely-spaced spikes right after $200$ ms on the time axis, and introduce spurious supports; (e)-(f) DuRIN-$1$ and DuRIN-$2$, on the other hand, distinguish between the closely-spaced spikes and exhibit superior support recovery, not introducing any spurious, low-amplitude supports.}\label{fig:1d_trace}
\end{figure*}

We validated the performance of DuRIN-$1$ and DuRIN-$2$ on $1000$ realizations of synthetic $1$-D traces, with a $30$-Hz Ricker wavelet, $1$-ms sampling interval, and amplitude increment $0.2$. Table \ref{table:1d_trace} shows the comparison of objective metrics for the proposed networks and the benchmark techniques. DuRIN-$1$ and DuRIN-$2$ recover the amplitudes with higher accuracy than the compared methods, quantified by the objective metrics CC, RRE, and SRER, with considerably superior support recovery indicated by PES. Figure \ref{fig:1d_trace} shows the corresponding illustration for a sample $1$-D seismic trace of the $1000$ test realizations. The figure illustrates two main advantages of the proposed networks. Both DuRIN-$1$ and DuRIN-$2$ resolve closely-spaced reflectors right after $200$ ms on the time axis in Figure \ref{fig:1d_trace} (e) and (f) and do not introduce spurious supports as the benchmark techniques. 

\begin{table}[t]
    \centering
    \caption{Objective metrics averaged over $1000$ test realizations of synthetic $1$-D seismic traces. The proposed DuRIN-$1$ and DuRIN-$2$ are superior in both amplitude and support recovery. The best performance is highlighted in \textbf{boldface}, while the second best is \underline{underlined}.}
    \label{table:1d_trace}
    \resizebox{0.85\columnwidth}{!}{
    \begin{tabular}{l||c|c|c|c}
        \toprule
        \bfseries Method & \bfseries CC & \bfseries RRE & \bfseries SRER & \bfseries PES\\
        \midrule
        BPI & ${0.6611}$ & ${0.5723}$ & ${3.4148}$ & ${0.9697}$\\
        FISTA & ${0.6734}$ & ${0.5489}$ & ${3.5554}$ & ${0.8309}$\\
        SBL-EM & ${0.6561}$ & ${0.6611}$ & $\mathbf{4.2666}$ & ${0.9697}$\\
        DuRIN-$1$ & $\mathbf{0.7259}$ & $\mathbf{0.4638}$ & ${4.0212}$ & $\underline{0.5309}$\\
        DuRIN-$2$ & $\underline{0.7237}$ & $\underline{0.4660}$ & $\underline{4.0230}$ & $\mathbf{0.5308}$\\
        \bottomrule
    \end{tabular}}
\end{table}

Table \ref{table:time} gives a comparison of the computational testing time over $100$, $200$, $500$, and $1000$ test realizations of synthetic $1$-D traces. DuRIN-$1$ and DuRIN-$2$ require lower computational time than the benchmark techniques, over two orders of magnitude lower than FISTA, the next best technique to our proposed networks. Lower computation times are significant for reflection seismic processing, where the size of datasets analyzed is large. We note that the lower computational testing times for DuRIN-$1$ and DuRIN-$2$ come at the expense of longer training times, where the models are trained on a large number of synthetic seismic traces.

\begin{table}[t]
    \centering
    \caption{Computational testing time (s) for $100$, $200$, $500$, and $1000$ test realizations of synthetic $1$-D seismic traces. DuRIN-$1$ and DuRIN-$2$ require significantly lower compute time during the inference phase as compared with the benchmark techniques. \\ R = number realizations.}
    \label{table:time}
    \resizebox{0.9\columnwidth}{!}{
    \begin{tabular}{l||c|c|c|c}
        \toprule
        \multirow{2}{*}{\bfseries Method} & \multicolumn{4}{c}{\bfseries Testing time ($\mathrm{s}$)}\\
        \cmidrule{2-5}
         & {\bfseries R} $=\mathbf{100}$ & {\bfseries R} $=\mathbf{200}$ & {\bfseries R} $=\mathbf{500}$ & {\bfseries R} $=\mathbf{1000}$ \\
        \midrule
        BPI & ${34.9307}$ & ${70.5120}$ & ${171.6517}$ & ${375.6346}$\\
        SBL-EM & ${202.5606}$ & ${400.1070}$ & ${947.3621}$ & ${1959.4553}$\\
        FISTA & ${7.1995}$ & ${13.1076}$ & ${34.1289}$ & ${64.9389}$\\
        DuRIN-$1$ & $\mathbf{0.0220}$ & $\underline{0.0467}$ & $\underline{0.0804}$ & $\underline{0.1434}$\\
        DuRIN-$2$ & $\underline{0.0276}$ & $\mathbf{0.0365}$ & $\mathbf{0.0732}$ & $\mathbf{0.1095}$\\
        \bottomrule
    \end{tabular}}
\end{table}

\begin{figure*}[t]
    \centering
    \includegraphics[width=0.55\linewidth]{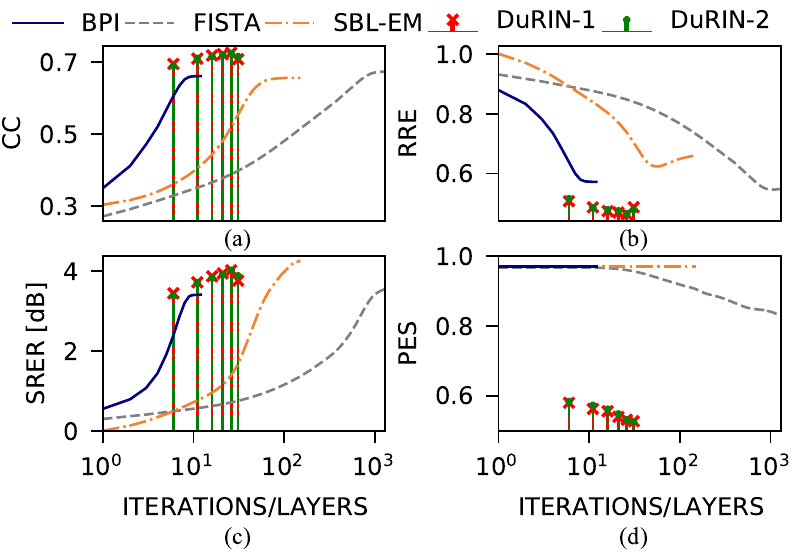}
    \caption{\protect\rgbsymbol~Objective metrics vs. number of iterations/layers for the five methods considered in the study. The stems represent layers ($6$, $11$, $16$, $21$, $26$, and $31$) for DuRIN-$1$ and DuRIN-$2$, and the lines represent iterations for BPI, FISTA, and SBL-EM. For (a) CC and (c) SRER, higher values indicate superior performance, while for (b) RRE and (d) PES, lower values indicate better performance.}\label{fig:iters_1d_trace}
\end{figure*}

Figure \ref{fig:iters_1d_trace} illustrates the computational advantage of the proposed DuRIN-$1$ and DuRIN-$2$ over the benchmark techniques in terms of layers/iterations. The $6$-layer DuRIN-$1$ and DuRIN-$2$ models outperform the benchmark techniques in terms of the four objective metrics considered in this study, with their performance further enhanced by adding $5$ layers, up to $26$ layers. We observe that the performance starts deteriorating after $31$ layers (Figure \ref{fig:iters_1d_trace}). We hypothesize that this deterioration could be attributed to the increase in the number of network parameters while keeping the size of the training dataset fixed, leading to overfitting. Future work could explore if increasing the training examples allows us to append more layers to enhance the networks' performance further rather than diminish it.

\subsubsection{Synthetic 2-D Wedge Models}\label{subsubsec:wedge}
We use synthetic wedge models to test the resolving capability of the proposed networks on thin beds \cite{hamlyn2014thin}. Wedge models usually consist of two interfaces, one horizontal and another inclined, with the wedge thinning to $0$ ms. Depending on the polarity of the reflection coefficients of the two interfaces being either the same or opposite, we have even or odd wedge models, respectively. Further, based on the polarity being negative or positive, we obtain four possible combinations of polarities of the reflection coefficients of the wedge models (NP, PN, NN, PP). In our experimental setup, we consider wedge models with $26$ seismic traces, with the separation between the two interfaces reducing from $50$ ms to $0$ ms in $2$ ms increments, and amplitudes of the reflection coefficients set to $\pm~0.5$. 

\begin{figure*}[t]
    \centering
    \includegraphics[width=0.6\linewidth]{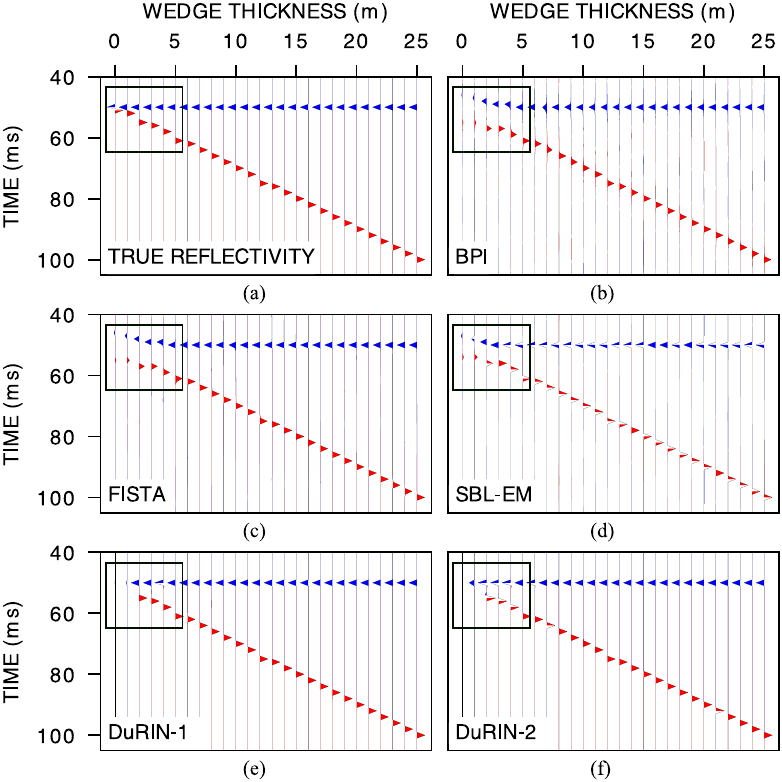}
    \caption{\protect\rgbsymbol~Results for a synthetic $2$-D odd wedge model. (a) True reflectivity; (b)-(d) Recovered reflectivity profiles from the benchmark techniques show a divergence from the ground-truth at wedge thickness $<5$ m, highlighted by the rectangle in black; (e)-(f) Reflectivity profiles recovered by DuRIN-$1$ and DuRIN-$2$ show a better resolution of reflectors below a wedge thickness of $5$ m.}\label{fig:wedge_np}
\end{figure*}

Table \ref{table:2D_np} and Figure \ref{fig:wedge_np} show the performance of the proposed networks in comparison with the benchmark techniques for an odd wedge model (NP), with negative reflection coefficients on the upper horizontal interface (N) and positive on the lower inclined interface (P). DuRIN-$1$ and DuRIN-$2$ outperform the other methods in terms of both amplitude and support recovery metrics (Table \ref{table:2D_np}). As highlighted in Figure \ref{fig:wedge_np}, BPI, FISTA, and SBL-EM fail to resolve the reflectors below $5$ m wedge thickness, which is below the tuning thickness of $13$ ms, corresponding to $6.5$ m in our experimental setup, for a $30$ Hz wavelet frequency \cite{chung1995frequency}. This low resolving capability also contributes to the poorer (higher) PES scores for these techniques (Table \ref{table:2D_np}). On the other hand, the proposed DuRIN-$1$ and DuRIN-$2$ maintain the lateral continuity of both interfaces even below the tuning thickness. 

\begin{table}[t]
    \caption{Metrics for a synthetic $2$-D odd (NP) wedge model. DuRIN-$1$ and DuRIN-$2$ outperform the benchmark techniques in terms of the amplitude and support recovery metrics considered. The best performance is highlighted in \textbf{boldface}. The second best scores are \underline{underlined}.}
    \label{table:2D_np}
    \centering
    \resizebox{0.85\columnwidth}{!}{
    \begin{tabular}{l||c|c|c|c}
        \toprule
        \bfseries Method & \bfseries CC & \bfseries RRE & \bfseries SRER & \bfseries PES\\
        \midrule
        BPI & ${0.8110}$ & ${0.2693}$ & ${12.4474}$ & ${0.9933}$\\
        FISTA & ${0.8108}$ & ${0.2583}$ & ${20.5075}$ & ${0.8846}$\\
        SBL-EM & ${0.8848}$ & ${0.1446}$ & ${34.1678}$ & ${0.9933}$\\
        DuRIN-$1$ & $\mathbf{0.9875}$ & $\mathbf{0.0622}$ & $\underline{35.7206}$ & $\underline{0.1795}$\\
        DuRIN-$2$ & $\underline{0.9774}$ & $\underline{0.0734}$ & $\mathbf{37.6269}$ & $\mathbf{0.1026}$\\
        \bottomrule
    \end{tabular}}
\end{table}

\begin{figure*}[ht]
    \centering
    \includegraphics[width=0.6\linewidth]{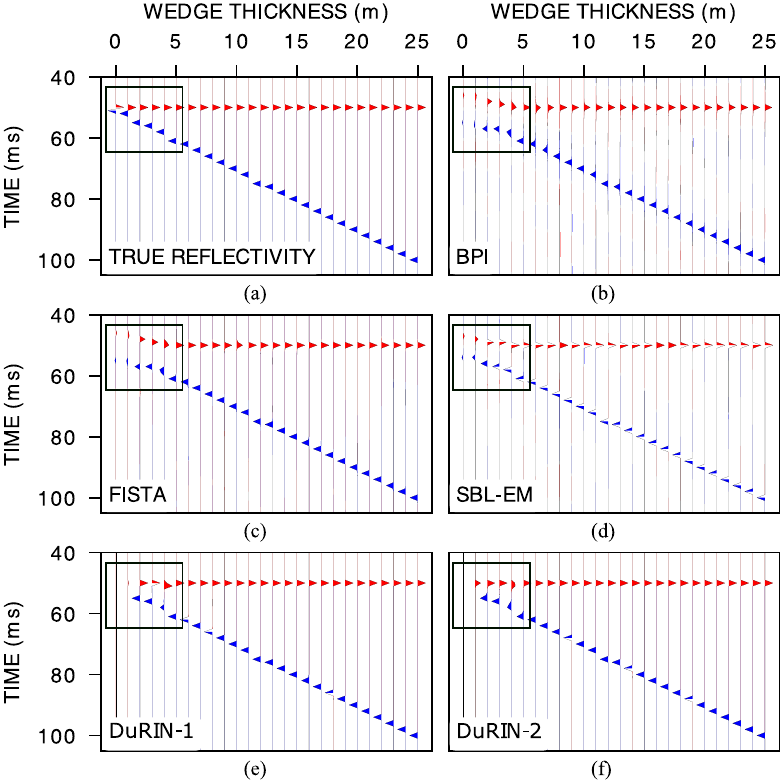}
    \caption{\protect\rgbsymbol~Results for recovered reflectivity from a synthetic $2$-D odd wedge model (PN). (a) True reflectivity; (b)-(d) Recovered reflectivity signatures from the benchmark methods show a distinct divergence from the ground-truth reflectivity (a) at wedge thickness $<~5$ m, highlighted by the rectangle in black; (e)-(f) Recovered reflectivity profiles from the proposed networks, namely, DuRIN-$1$ and DuRIN-$2$ resolve the reflectors below a wedge thickness of $5$ m.}\label{fig:wedge_pn}
\end{figure*}

Table \ref{table:2D_pn} and Figure \ref{fig:wedge_pn} show the results for an odd wedge model (PN) with positive polarity of reflection coefficients on the upper, horizontal interface (P) and negative polarity on the lower, inclined interface (N). Table \ref{table:2D_pn} shows the proposed DuRIN-$1$ and DuRIN-$2$ outperforming the benchmark techniques, namely, BPI, FISTA, and SBL-EM, in terms of the amplitude and support recovery metrics considered in this study. Similar to the NP wedge model, and as highlighted in Figure \ref{fig:wedge_pn}, the benchmark techniques do not accurately resolve the reflectors below a wedge thickness of $5$ m, whereas DuRIN-$1$ and DuRIN-$2$ preserve lateral continuity of both the interfaces of the wedge model below tuning thickness. 
\begin{table}[t]
    \caption{Results for a synthetic $2$-D odd (PN) wedge model. The proposed networks, namely, DuRIN-$1$ and DuRIN-$2$ are superior in both amplitude and support recovery accuracy.}
    \label{table:2D_pn}
    \centering
    \resizebox{0.85\columnwidth}{!}{
    \begin{tabular}{l||c|c|c|c}
        \toprule
        \bfseries Method & \bfseries CC & \bfseries RRE & \bfseries SRER & \bfseries PES\\
        \midrule
        BPI & ${0.8369}$ & ${0.2265}$ & ${12.5753}$ & ${0.9933}$\\
        FISTA & ${0.8387}$ & ${0.2123}$ & ${21.5266}$ & ${0.8832}$\\
        SBL-EM & ${0.8469}$ & ${0.2106}$ & ${32.4996}$ & ${0.9933}$\\
        DuRIN-$1$ & $\mathbf{0.9774}$ & $\mathbf{0.0795}$ & $\underline{34.4034}$ & $\underline{0.1923}$\\
        DuRIN-$2$ & $\underline{0.9577}$ & $\underline{0.1074}$ & $\mathbf{36.0748}$ & $\mathbf{0.1256}$\\
        \bottomrule
    \end{tabular}}
\end{table}

\begin{figure}[t]
    \centering
    \includegraphics[width=\linewidth]{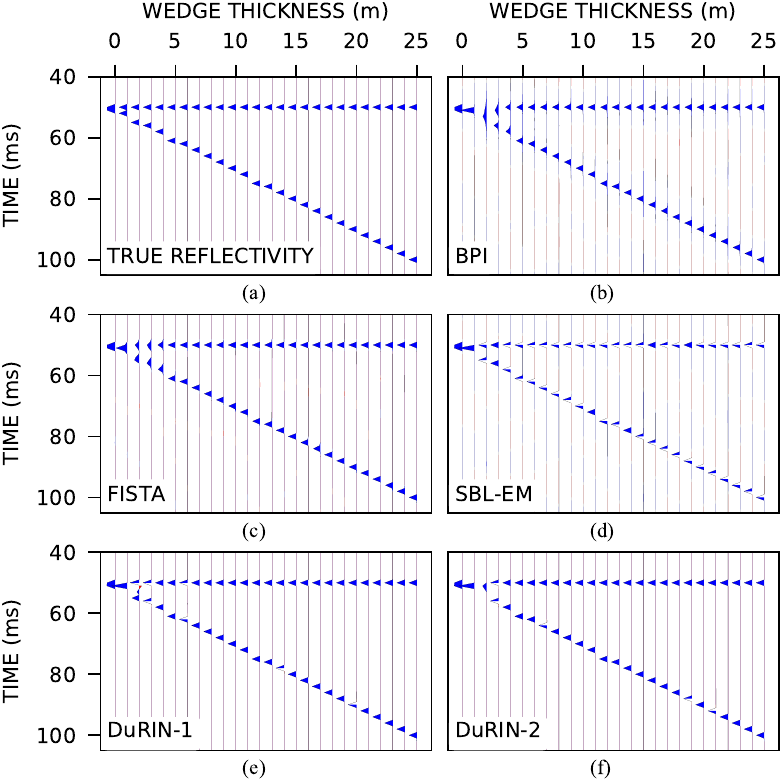}
    \caption{\protect\rgbsymbol~Results for recovered reflectivity from a synthetic $2$-D even wedge model (NN). (a) True reflectivity; (b)-(f) Recovered reflectivity profiles. The proposed networks and the benchmark techniques resolve the even wedge model well for wedge thickness $>~3$ m. DuRIN-$1$ (e), along with SBL-EM (d), also resolves the two reflectors of trace $3$ (corresponding to wedge thickness $2$ m).}\label{fig:wedge_nn}
\end{figure}

\begin{figure}[ht]
    \centering
    \includegraphics[width=\linewidth]{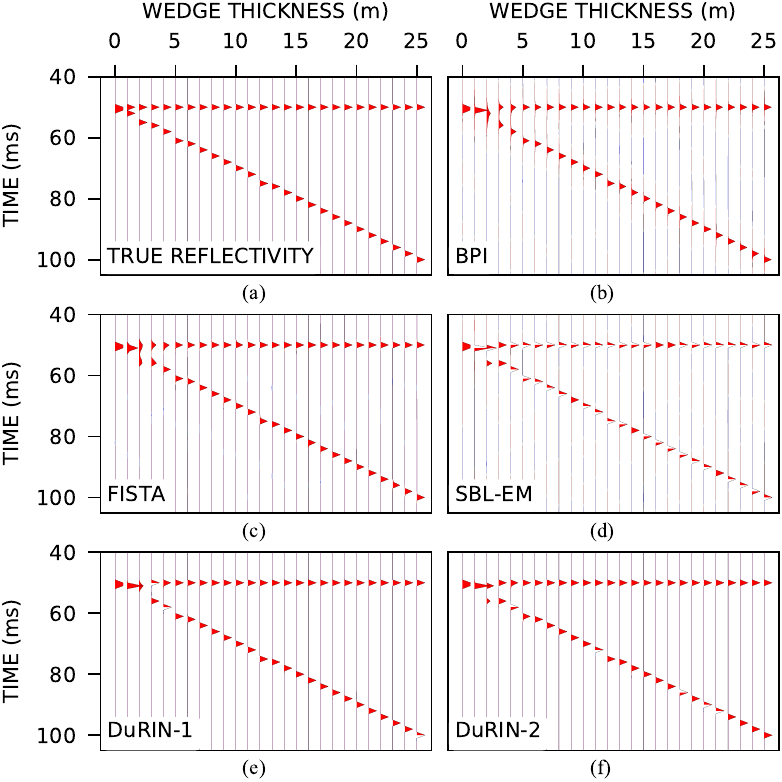}
    \caption{\protect\rgbsymbol~Results for an even wedge model (PP). (a) True; and (b)-(f) Recovered reflectivity. BPI, FISTA, and DuRIN-$1$ resolve reflectors for wedge thickness $>~3$ m, while SBL-EM and DuRIN-$2$ also resolve reflectors at $2$ m.}\label{fig:wedge_pp}
\end{figure}

For the even wedge models, i.e., NN (Table \ref{table:2D_nn} and Figure \ref{fig:wedge_nn}) and PP (Table \ref{table:2D_pp} and Figure \ref{fig:wedge_pp}), all techniques exhibit comparable performance in terms of the amplitude metrics of CC and RRE, and resolve reflectors below tuning thickness without any divergence from the ground-truth reflector locations. The proposed networks, namely, DuRIN-$1$ and DuRIN-$2$ outperform the benchmark techniques in SRER and PES scores.
\begin{table}[ht]
    \centering
    \caption{Results for an even wedge model (NN). 
    DuRIN-$1$ and DuRIN-$2$ are superior in terms of SRER and PES.}
    \label{table:2D_nn}
    \resizebox{0.85\linewidth}{!}{
    \begin{tabular}{l||c|c|c|c}
        \toprule
        \bfseries Method & \bfseries CC & \bfseries RRE & \bfseries SRER & \bfseries PES\\
        \midrule
        BPI & ${0.9215}$ & ${0.1733}$ & ${15.1795}$ & ${0.9933}$\\
        FISTA & ${0.9572}$ & $\mathbf{0.0916}$ & ${22.1490}$ & ${0.8643}$\\
        SBL-EM & $\mathbf{0.9616}$ & $\underline{0.1120}$ & ${38.1103}$ & ${0.9933}$\\
        DuRIN-$1$ & ${0.9235}$ & ${0.1568}$ & $\mathbf{38.8751}$ & $\underline{0.1603}$\\
        DuRIN-$2$ & $\underline{0.9599}$ & ${0.1163}$ & $\underline{38.2828}$ & $\mathbf{0.1346}$\\
        \bottomrule
    \end{tabular}}
\end{table}
\begin{table}[ht]
    \caption{Results for an even wedge model (PP). 
    While FISTA has a better CC and RRE score, DuRIN-$1$ and DuRIN-$2$ outperform the benchmark techniques in terms of SRER and PES.}
    \label{table:2D_pp}
    \centering
    \resizebox{0.85\linewidth}{!}{
    \begin{tabular}{l||c|c|c|c}
        \toprule
        \bfseries Method & \bfseries CC & \bfseries RRE & \bfseries SRER & \bfseries PES\\
        \midrule
        BPI & ${0.9303}$ & ${0.1524}$ & ${15.5630}$ & ${0.9933}$\\
        FISTA & $\mathbf{0.9548}$ & $\mathbf{0.0905}$ & ${23.1071}$ & ${0.8572}$\\
        SBL-EM & ${0.9232}$ & ${0.1882}$ & ${36.2027}$ & ${0.9933}$\\
        DuRIN-$1$ & ${0.9238}$ & ${0.1635}$ & $\mathbf{36.9497}$ & $\underline{0.1923}$\\
        DuRIN-$2$ & $\underline{0.9295}$ & $\underline{0.1447}$ & $\underline{36.8566}$ & $\mathbf{0.1410}$\\
        \bottomrule
    \end{tabular}}
\end{table}

\subsection{Testing Phase: Marmousi2 Model}\label{subsec:marmousi2}
The Marmousi$2$ model \cite{martin2006marmousi2} (width $\times$ depth: $17$ km $\times~3.5$ km) is an expansion of the original Marmousi model \cite{versteeg1994marmousi} (width $\times$ depth: $9.2$ km $\times~3$ km depth). The model provides a simulated $2$-D collection of seismic traces, and is widely used in reflection seismic processing for the calibration of techniques in structurally complex settings. The model has a $2$ ms sampling interval, with traces at an interval of $12.5$ m. We obtained the ground-truth reflectivity profiles $({\boldsymbol{x}})$ from the P-wave velocity and density models \eqref{rhov}, and convolved them with a $30$ Hz Ricker wavelet to generate the input data $({\boldsymbol{y}})$, with measurement SNR of $20$ dB to test the robustness of the proposed networks to noisy conditions. 

\begin{table}
    \caption{Results for a portion of the Marmousi$2$ model, corresponding to Table \ref{table:mm_mute}, without muting low-amplitude spikes. DuRIN-$1$ and DuRIN-$2$ have higher CC and RRE. The lower PES scores for BPI and SBL-EM are due to the spurious supports introduced by these techniques coinciding with the low-amplitude spikes. The best performance is highlighted in \textbf{boldface}; the second best is \underline{underlined}.} 
    \label{table:mm_unmute}
    \centering
    \resizebox{0.85\linewidth}{!}{
    \begin{tabular}{l||c|c|c|c}
        \toprule
        \bfseries Method & \bfseries CC & \bfseries RRE & \bfseries SRER & \bfseries PES\\
        \midrule
        BPI & ${0.9714}$ & ${0.0474}$ & ${22.2817}$ & $\mathbf{0.1751}$\\
        FISTA & ${0.9716}$ & ${0.0466}$ & $\underline{23.1565}$ & $\underline{0.8879}$\\
        SBL-EM & ${0.9808}$ & ${0.0305}$ & $\mathbf{26.3986}$ & $\mathbf{0.1751}$\\
        DuRIN-$1$ & $\mathbf{0.9857}$ & $\mathbf{0.0243}$ & ${22.5103}$ & ${0.9716}$\\
        DuRIN-$2$ & $\underline{0.9820}$ & $\underline{0.0302}$ & ${22.1734}$ & ${0.9702}$\\
        \bottomrule
    \end{tabular}}
\end{table}

\begin{table}[t]
    \caption{Objective metrics for a portion of the Marmousi$2$ model \cite{martin2006marmousi2}. DuRIN-$1$ and DuRIN-$2$ exhibit superior amplitude recovery than the benchmark techniques in terms of CC and RRE, and outperform in support recovery in terms of PES.}
    \label{table:mm_mute}
    \centering
    \resizebox{0.85\columnwidth}{!}{
    \begin{tabular}{l||c|c|c|c}
        \toprule
        \bfseries Method & \bfseries CC & \bfseries RRE & \bfseries SRER & \bfseries PES\\
        \midrule
        BPI & ${0.9717}$ & ${0.0465}$ & ${23.4547}$ & ${0.9778}$\\
        FISTA & ${0.9720}$ & ${0.0456}$ & $\underline{25.1384}$ & ${0.7762}$\\
        SBL-EM & ${0.9811}$ & ${0.0300}$ & $\mathbf{30.5194}$ & ${0.9778}$\\
        DuRIN-$1$ & $\mathbf{0.9860}$ & $\mathbf{0.0238}$ & ${24.5692}$ & $\mathbf{0.2175}$\\
        DuRIN-$2$ & $\underline{0.9822}$ & $\underline{0.0298}$ & ${23.8769}$ & $\underline{0.2438}$\\
        \bottomrule
    \end{tabular}}
\end{table}%

\begin{figure*}[!ht]
    \centering
    \includegraphics[width=0.65\linewidth]{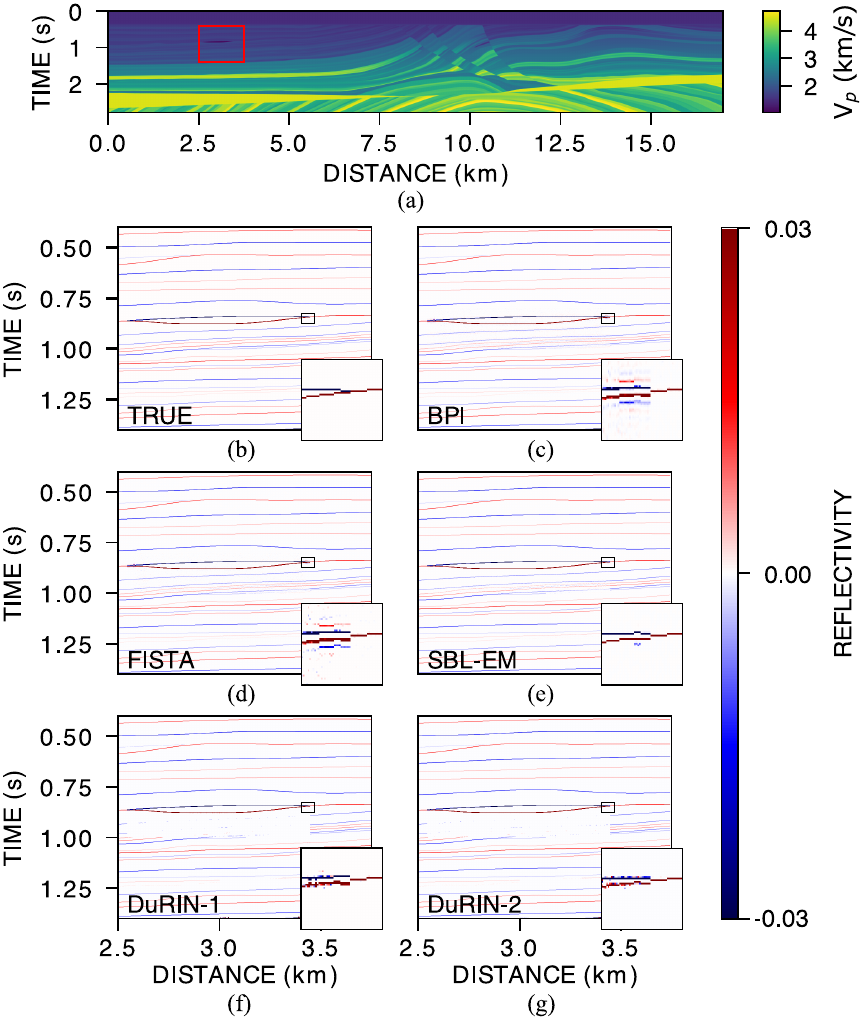}
    \caption{\protect\rgbsymbol~Results for a portion of the Marmousi$2$ synthetic model, corresponding to Table \ref{table:mm_mute} and marked by a red rectangle in (a) The P-wave velocity profile for the complete model; (b) The true reflectivity; (c)-(g) recovered $2$-D reflectivity profiles from the methods compared in this study, with insets plots showing the zoomed-in portions of the selected region marked by black rectangles. Insets show that (c) BPI and (d) FISTA introduce spurious reflection coefficients around the true interfaces, while (e) SBL-EM, (f) DuRIN-$1$, and (g) DuRIN-$2$ preserve the lateral continuity better.}\label{fig:marmousi}
\end{figure*}

Here, we present results from a region of the Marmousi$2$ model that contains a gas-charged sand channel \cite{martin2006marmousi2}. Table \ref{table:mm_unmute} shows a comparison of the performance of the proposed DuRIN-$1$ and DuRIN-$2$ with that of the benchmark techniques for the portion containing the gas-charged sand channel. The Marmousi$2$ model \cite{martin2006marmousi2} has a large number of low-amplitude reflections. Additionally, in Figure \ref{fig:1d_trace}, we observe that BPI and SBL-EM introduce spurious reflections. We attribute the low PES scores observed for BPI and SBL-EM in Table \ref{table:mm_unmute} to such spurious supports complementing the low-amplitude reflections in the Marmousi$2$ model. We re-evaluate the objective metrics after muting reflections with amplitudes $<~1\%$ of the absolute of the maximum amplitude, the results for which are reported in Table \ref{table:mm_mute} and Figure \ref{fig:marmousi}. The re-evaluation did not affect the CC, RRE, and SRER scores adversely, but, as the PES is computed over significant reflections, it is higher (worse) for BPI and SBL-EM, and lower (better) for DuRIN-$1$ and DuRIN-$2$. The objective metrics reported in Table \ref{table:mm_mute} show superior amplitude and support recovery accuracy of the proposed networks over the benchmark methods. Figure \ref{fig:marmousi} shows that DuRIN-$1$ and DuRIN-$2$, along with SBL-EM, preserve the lateral continuity of interfaces, highlighted in the insets at the edge of the sand channel. The insets also show that BPI and FISTA introduce spurious false interfaces due to the interference of multiple events. DuRIN-$1$ and DuRIN-$2$ show limited recovery of the low-amplitude reflection coefficients right below the gas-charged sand channel (Figure \ref{fig:marmousi} (f) and (g)), which could be investigated in future work.

\subsection{Testing Phase: Real Data}\label{subsec:real_data}
To validate the proposed networks on real data, we use a $3$-D seismic volume from the Penobscot $3$D survey off the coast of Nova Scotia, Canada \cite{penobscot3d}. We present results from a smaller region of the $3$-D\\ volume, with $201$ Inlines ($1150$-$1350$) and $121$ Xlines ($1040$-$1160$), a region that includes the two wells of this survey (wells L-$30$ and B-$41$, \cite{bianco2014geophysical}). Along the time axis, the region contains $800$ samples between the time interval of $0$ to $3196$ ms, with a $4$ ms sampling interval, and a $25$ Hz Ricker wavelet fits the data well \cite{bianco2014geophysical}.

\begin{figure*}[t]
    \centering
    \includegraphics[width=0.65\linewidth]{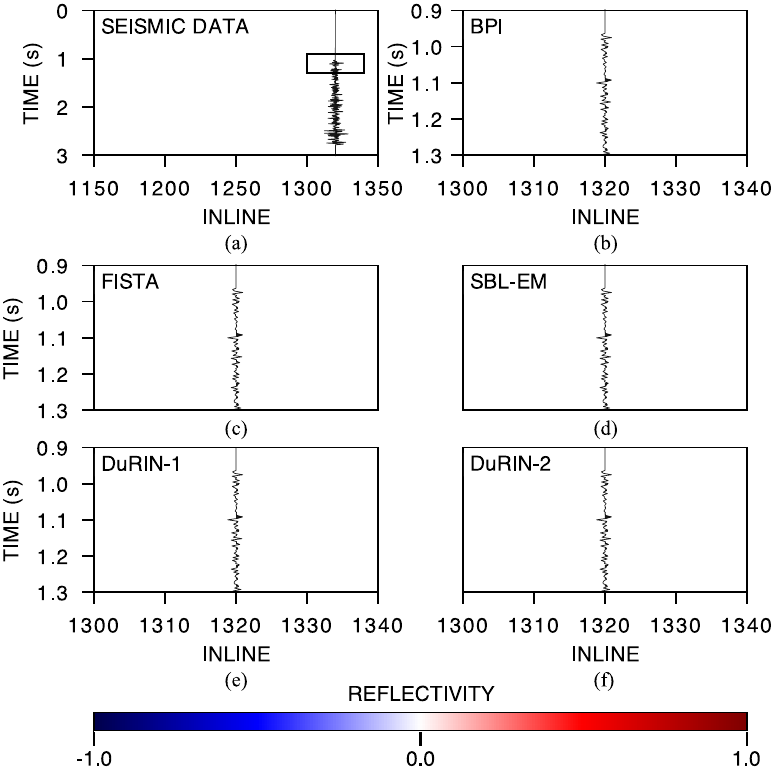}
    \caption{\protect\rgbsymbol~(a) Observed seismic data; (b)-(f) predicted $2$-D reflectivity profiles for the region marked by a black rectangle in (a), for Xline $1155$ of the Penobscot $3$D survey \cite{penobscot3d}. The overlaid waveforms in black are the seismic (in (a)) and reflectivity (in (b)-(f)), respectively, computed from the sonic logs of well L-$30$ \cite{bianco2014geophysical}. BPI (b) and FISTA (c) provide a smooth profile without many details, while SBL-EM (d), DuRIN-$1$ (e) and DuRIN-$2$ (f) generate a detailed reflectivity profile.}\label{fig:real}
\end{figure*}

Figure \ref{fig:real} shows the observed seismic data and recovered reflectivity profiles for Xline $1155$ of the survey, and overlaid in black, the seismic and reflectivity profiles, respectively, computed from the sonic logs of well L-$30$ \cite{bianco2014geophysical}. The inverted reflectivity profiles for BPI and FISTA are smooth and lack detail. On the other hand, the predicted reflectivity profiles generated using SBL-EM, DuRIN-$1$ and DuRIN-$2$ provide more details for characterizing the subsurface. Additionally, DuRIN-$1$ and DuRIN-$2$ also resolve closely-spaced interfaces better and preserve the lateral continuity, evident from the interfaces around $1.1$ s on the time axis (Figure \ref{fig:real} (e) and (f)). These results demonstrate the capability of the proposed networks, namely, DuRIN-$1$ and DuRIN-$2$, on real data from the field. We note that the models are trained on synthesized seismic traces before being tested on the real data.

\section{Conclusions}\label{sec:conclusions}
We developed a nonconvex optimization cost for reflectivity inversion based on a weighted counterpart of the minimax concave penalty. We also developed the deep-unfolded reflectivity inversion network (DuRIN) for solving the problem under consideration. Experimental validation on synthetic, simulated, and real data demonstrated the superior resolving capability of DuRIN, especially in support recovery (represented by PES), which is crucial for characterizing the subsurface. DuRIN also preserves lateral continuity of interfaces despite being a single-trace approach. However, the trade-off between the number of layers of the network and training examples, and the suboptimal recovery of low-amplitude reflection coefficients require further attention. One could also consider data-driven prior learning \cite{bergen2019machine} based on a multi-penalty formulation \cite{jawali2020cornet} for solving the reflectivity inversion problem. 

\section*{Acknowledgment}
This work is supported by the Ministry of Earth Sciences, Government of India; Centre of Excellence in Advanced Mechanics of Materials, Indian Institute of Science (IISc), Bangalore; and Science and Engineering Research Board (SERB), India. The authors would like to thank Jishnu Sadasivan for fruitful technical discussions.

\bibliography{durin_arxiv}
\bibliographystyle{IEEEtran}









\end{document}